\documentclass[english,aps,prl,superscriptaddrkess,floatfix,notitlepage,reprint,show pacs]{revtex4-2}
\usepackage[T1]{fontenc}
\usepackage[utf8]{inputenc}
\usepackage{physics}
\usepackage{natbib}
\usepackage{amsthm}
\usepackage{float}
\usepackage{amssymb}
\usepackage{dsfont}
\usepackage{amsmath}
\usepackage{bm}
\usepackage{sublabel}
\usepackage{latexsym}
\usepackage{sidecap}
\usepackage{placeins}
\usepackage{url}
\makeatletter
\theoremstyle{plain}

\theoremstyle{plain}

\theoremstyle{plain}
\newtheorem*{prop*}{\protect\propositionname}

\usepackage{braket}
\usepackage{txfonts}
\usepackage{pifont}
\usepackage{graphicx}
\usepackage[dvipsnames]{xcolor}
\usepackage{hyperref}
\usepackage{cleveref}
\hypersetup{
    colorlinks=true,
    linkcolor=Red,       
    citecolor=blue,      
    urlcolor=cyan      
}
\usepackage{orcidlink}
\usepackage{tikz}
\usetikzlibrary{patterns,decorations.text,decorations.pathreplacing,decorations.pathmorphing}
\usepackage{caption}
\usepackage{subcaption}
\usepackage{rotating}
\usepackage{lipsum}

\date{\today}
\newcommand{\beq}{\begin{equation}}
\newcommand{\eeq}{\end{equation}}
\newcommand{\beqa}{\begin{eqnarray}}
\newcommand{\eeqa}{\end{eqnarray}}

\newcommand{\blackbullet}{%
  \protect\tikz[baseline=-0.6ex]
  \fill[black] (0,0) circle (0.35em);}

\newcommand{\red}{\protect\tikz[baseline=-0.5ex]\draw[red] (0,0)--(0.5,0);}
\newcommand{\blue}{\protect\tikz[baseline=-0.5ex]\draw[blue] (0,0)--(0.5,0);}
\newcommand{\green}{\protect\tikz[baseline=-0.5ex]\draw[green] (0,0)--(0.5,0);}
\newcommand{\black}{\protect\tikz[baseline=-0.5ex]\draw[black] (0,0)--(0.5,0);}
\newcommand{\cyan}{\protect\tikz[baseline=-0.5ex]\draw[cyan] (0,0)--(0.5,0);}

\newcommand{\magenta}{\protect\tikz[baseline=-0.5ex]\draw[magenta] (0,0)--(0.5,0);}

\newcommand{\reddashed}{\protect\tikz[baseline=-0.5ex]\draw[red,dashed] (0,0)--(0.5,0);}

\begin{document}
\title{Fluctuation Spectra and Response Function of Coupled Atomic and Molecular BECs}
\author{Avinaba Mukherjee \orcidlink{0009-0000-3765-6466}}
\thanks{\href{mailto:avinaba.mukherjee@rediffmail.com}{avinaba.mukherjee@rediffmail.com}}
\author{Raka Dasgupta \orcidlink{0000-0003-2148-4641}}
\thanks{\href{mailto:rdphy@caluniv.ac.in}{rdphy@caluniv.ac.in}}
\address{Department of Physics, University of Calcutta, $92$ A. P. C. Road, Kolkata $700009$, India}
\begin{abstract}

We investigate out-of-equilibrium properties of atomic–molecular Bose–Einstein condensates coupled through a Feshbach resonance, with the Feshbach coupling and detuning subject to Gaussian white noise. Using a bosonic Josephson-junction framework and a Bloch-sphere description, we examine the interplay of detuning, coherence, and noise governing the system dynamics. Coupling and detuning noise produce distinct fluctuation spectra, featuring both Feshbach-resonant and symmetric off-resonant peaks. We characterize the dispersive and absorptive response of the atom–dimer system under periodic driving. The atom–molecule hybridization at the Feshbach resonance maximizes the linewidth and minimizes both the effective temperature and the phase difference between the driving field and the system. This leads to an optimized power utilization and quality factor.

\end{abstract}
\maketitle
%\begin{multicols}{2}
\section{Introduction}
Coupled atomic (A) - molecular (M)  Bose-Einstein Condensates (BECs) form a rich platform exhibiting diverse quantum-statistical phenomena \cite{drummond1998coherent,timmermans1999rarified,javanainen1999coherent,heinzen2000superchemistry}. Here, two atoms combine into a molecule via a Feshbach resonance \cite{chin2010feshbach}, mapping the system onto a bosonic Josephson junction \cite{BJJ9,BJJ18,BJJ20}, with atom--molecule coupling analogous to tunneling. The A-BEC and M-BEC are separated by a tunable formation threshold, similar to a Josephson barrier, which controls the system dynamics. External control through detuning can give rise to bistability, with regimes dominated either by atoms or by molecules \cite{kohler2006production}. The dissipative bosonic Josephson junction has been extensively investigated as an open quantum system   \cite{khripkov2011quantum,Linblad_master_equation,witthaut2009dissipation}.\\
For coherent evolution, the Bloch vector undergoes deterministic precession around an effective field, where detuning controls both the axis of rotation and the precessional frequency \cite{kittel2005solid}. Additionally, in the presence of diffusion, the Bloch vector exhibits diffusive spreading arising from environmental interactions (noise) and external modulation (detuning) \cite{saha2023phase}. The resulting stochastic dynamics can therefore be quantified by the diffusion coefficients associated with the principal axes.
Detuning measures the mismatch between the system’s intrinsic transition frequency and the external driving frequency \cite{Taylor2005}, and it plays a key role in shaping this diffusion process.\\
The presence of noise in the atom-dimer two-state model can render the system dynamics analogous to those of a Brownian oscillator \cite{P_many}. The static mobility and diffusion coefficients determine the effective temperatures of the dynamical resonance mode \cite{p323}. By establishing a correspondence between the periodically driven atom-dimer system and a damped harmonic oscillator, we analyze the power absorption \cite{P14}, the quality factor \cite{P261,P129}, and the full width at half maximum (FWHM) \cite{P304} of the response function \cite{P475}, focusing on their dependence on the Feshbach detuning.

In out-of-equilibrium systems, memory of the initial state can persist due to long relaxation times \cite{P563}, i.e., when the system is weakly coupled to the environment \cite{dutta2025introduction}. The noise intensity promotes transitions between the two levels; thus, all these quantities associated with the power spectrum are influenced by stochastic variable, noise \cite{wellens2004stochastic}.

In this manuscript, we investigate Feshbach-coupled atomic-molecular condensates in which both the Feshbach coupling and detuning are independently subjected to Gaussian white noise. We focus on the non-equilibrium dynamics of coherence and polarization fluctuations around a stable equilibrium configuration, examining the real-time evolution of both the coherence and the atom-molecule population imbalance.

The interplay between detuning and noise gives rise to a fluctuation spectrum in which noise significantly enhances the diffusive dynamics. Furthermore, when the system is under the influence of a periodic external field, its response depends on the difference between the modulation frequency and the intrinsic fluctuation frequency of the Bloch components of this effective two-level atom-dimer system. As this frequency mismatch decreases, a dynamical resonance emerges, manifested by enhanced mobility and an increased population transfer between the two states.

The paper is organized as follows. Sec. \ref{2nd section} presents the formal description of the two-level system. In Sec. \ref{main mechanism}, we examine the fluctuation spectra of coherence and imbalance between these two species as functions of the Feshbach detuning. Sec. \ref{auto correlation time} is devoted to the analysis of mobility and the extraction of the effective temperature from the static mobility. In Sec. \ref{feature}, we explore several key characteristics of the system, including power absorption, the quality factor, and FWHM of the response function, highlighting their dependence on the tunable Feshbach detuning. Finally, we summarize our work in Sec. \ref{conclusion}.
\section{Model Hamiltonian and Dynamical Equations}\label{2nd section}
Sec. \ref{TSM} introduces the two-state model, while Sec. \ref{inclusion} addresses the inclusion of noise.
\subsection{Two state model}\label{TSM}
We consider a system where pairs of bosonic atoms coherently form  bosonic molecules through a Feshbach resonance \cite{AMBEC,AMBEC2,donley2002atom}. The process is described by a two channel model \cite{review}. When the energy levels of two atoms coincide, resonant coupling can occur, leading to the formation of a bosonic dimer that is energetically favourable \cite{pitaevskii2016bec}. The energy offset between the A-BEC and M-BEC, denoted by $\epsilon_b$, can be controlled via an external magnetic field. The dynamics of the coupled atom-molecule system is described by the Hamiltonian:
\cite{similar_hamiltonian1,similar_hamiltonian2,similar_hamiltonian3,similar_hamiltonian4}.\\
\begin{equation}
\label{hamiltonian}
    \begin{split}
        \hat{H}= & \frac{u_1}{2V} \hat{a}^\dagger \hat{a}^\dagger \hat{a}\hat{a}+ \frac{u_2}{2V} \hat{b}^\dagger \hat{b}^\dagger \hat{b}\hat{b}+ \frac{u_3}{V} \hat{a}^\dagger \hat{b}^\dagger \hat{b}\hat{a}\\ &+ \frac{g}{\sqrt{V}} (\hat{a}^\dagger \hat{a}^\dagger \hat{b}+ \hat{b}^\dagger \hat{a}\hat{a}) +\epsilon_b  \hat{b}^\dagger \hat{b}
    \end{split}
\end{equation}
Here, $\hat{a}^\dagger$ ($\hat{a}$) and $\hat{b}^\dagger$ ($\hat{b}$) are creation (annihilation) operators for atoms and molecules. Interaction parameters are $u_1$ (atom--atom), $u_2$ (molecule--molecule), and $u_3$ (atom--molecule), while $g$ represents the Feshbach coupling between atomic and molecular states; $V$ is the quantization volume. Both the A-BEC and M-BEC are here treated as single-mode condensates, justified  where spatial fluctuations and thermal excitations are neglected. This mirrors the single-order-parameter Gross–Pitaevskii framework \cite{P-527} (which successfully captures experimental trends at the qualitative \cite{savage2003bose,albiez2005direct} level), as well as the two-mode descriptions used for noisy double-well BEC systems \cite{gati2006primary}.\\

For visualization, we use a Bloch vector representation analogous to spin systems, mapping the fully molecular and fully atomic states to the North and South poles of the Bloch sphere. Unlike true spins, however, the Bloch vector components here do not obey SU(2) algebra \cite{bloch_vector4, fermion_number2, fermion_number4}.
We define the Bloch vector (Schwinger pseudo spin) operators as \cite{P70,P189,bloch4,commutator,quantumgas}:
$\hat{L}_x= \sqrt{2}(\hat{a}^\dagger \hat{a}^\dagger \hat{b} + \hat{b}^\dagger \hat{a} \hat{a})/N^{3/2}$, $\hat{L}_y=\sqrt{2}i(\hat{a}^\dagger \hat{a}^\dagger \hat{b} - \hat{b}^\dagger \hat{a} \hat{a})/N^{3/2}$, $\hat{L}_z=(2\hat{b}^\dagger \hat{b}-\hat{a}^\dagger \hat{a})/N$, and $N=2\hat{b}^\dagger \hat{b}+\hat{a}^\dagger \hat{a}$  

Here, $\hat{L}_x$ and $\hat{L}_y$ represent the real and imaginary parts of the atom molecule coherence, while $\hat{L}_z$ gives the population imbalance \cite{cui2012atom}. The total particles number is $N$, and the key commutation relations of the Bloch vector components, relevant to the system dynamics, are given in \cite{bloch4,liu2010shapiro,quantumgas}. For double well condensates \cite{Linblad_master_equation}, the Bloch components form a closed $\mathrm{SU}(2)$ algebra \cite{stochastic_bosonic_josephson_junction}, whereas in the two mode atom dimer model they obey $\mathrm{SU}(1,1)$ \cite{khripkov2011quantum}.\\
Defining the scaled parameters $U_1 = N u_1 / V$, $U_2 = N u_2 / V$, $U_3 = N u_3 / V$, and $\tilde{g} = g\sqrt{N/V}$, Eq. (\ref{hamiltonian}) with noise terms can be expressed in the large-$N$ limit in terms of $\hat{L}_i$ as
 \begin{equation}
\label{hamiltonian_large_N}
\begin{split}
\hat{\mathcal{H}} = \frac{\hat{H}}{N} 
= & \frac{U_1}{8}(\hat{L}_z - 1)^2
+ \frac{U_2}{32}(\hat{L}_z + 1)^2 \\
& - \frac{U_3}{8}(\hat{L}_z^2 - 1)
+ \frac{\tilde{g}}{\sqrt{2}} \hat{L}_x
+ \frac{\epsilon_b}{4}(\hat{L}_z + 1)
\end{split}
\end{equation}

The Bloch vector picture thus serves purely as a  visualization tool, and for analyzing the dynamics in terms of physically meaningful variables.\\

\subsection{Inclusion of noise}\label{inclusion}
 We introduce zero-mean, delta-correlated stochastic perturbations to the modified coupling strength, $\tilde{g}$ and detuning $\epsilon_b$ \cite{noise_property}. The condensed fraction forms the system, while thermal atoms constitute a bath \cite{anglin1997cold,ruostekoski1998bose}, justifying a Gaussian white-noise approximation \cite{burt1997coherence}. Coupling noise ($\gamma_x$) arises from decoherence due to elastic collisions between the thermal cloud and the BECs \cite{witthaut2008dissipation}, whereas detuning noise ($\gamma_z$) originates from magnetic field fluctuations near the Feshbach resonance \cite{bloch_vector4} and thermal fluctuations within the BEC \cite{saha2023phase}. The large $N$ limit is well justified, as experiments typically involve $10^5$ to $10^6$ atoms in magneto-optical traps \cite{strecker2003conversion,burt1997coherence}. Thus, the model captures fluctuations and decoherence but excludes particle loss. The system remains closed, with a conserved total particle number ($\dot{N}=0$).
 
 \begin{equation}
\label{hamiltonian_large_N}
\begin{split}
\hat{\mathcal{H}} = \frac{\hat{H}}{N} 
= & \frac{U_1}{8}(\hat{L}_z - 1)^2
+ \frac{U_2}{32}(\hat{L}_z + 1)^2 \\
& - \frac{U_3}{8}(\hat{L}_z^2 - 1)
+ \frac{\tilde{g} + \eta_x}{\sqrt{2}} \hat{L}_x
+ \frac{\epsilon_b + \eta_z}{4}(\hat{L}_z + 1)
\end{split}
\end{equation}

Here, $\eta_x$ and $\eta_z$ denote stochastic fluctuations in the Feshbach coupling and the detuning, respectively. These noise contributions are modeled as
$\langle\eta_i(t)\rangle =\langle \mathrm{d}w_i/\mathrm{d}t\rangle$, since the average contribution of the collisions vanishes \cite{P175} where $w_i(t)$ ($i \in \{x,z\}$) are independent Wiener processes. Their increments satisfy
$\langle \mathrm{d}w_i\mathrm{d}w_j \rangle = \gamma_i\delta_{ij}\mathrm{d}t/2$ \cite{stochastic_bosonic_josephson_junction}.\\

In Sec. \ref{main mechanism}, we first analyze fluctuations along the principal axes via the drift and diffusion matrices, and then study how the fluctuation matrix elements depend on the Feshbach detuning, $\epsilon_b$.

\section{Dynamics of Bloch vector's components}\label{main mechanism}

This two-state model has two equilibrium points, $(0,0,1)$ and $(0,0,-1/3)$, of which the latter is relatively more stable \cite{quantumgas2}. By taking the expectation values of the Bloch-vector components, averaging over the noise, and linearizing the dynamics about the equilibrium point $(0,0,-1/3)$ while neglecting constant terms, we obtain the linearized system. Details of the derivation are given in Appendix \ref{appendix MF}, where
$k=\left(4U_1/3\hbar-U_2/6\hbar-U_3/3\hbar-\epsilon_b/\hbar\right)$.\\

\begin{subequations}
\label{noise}
\begin{equation}
\label{noisex}
 \delta \dot{s}_x=k \delta s_y-\frac{\gamma_z}{2} \delta s_x    
\end{equation}
\begin{equation}
\label{noisey}
\delta\dot{s}_y=-k \delta s_x-2\sqrt{2}g \delta s_z-(4\gamma_x+\frac{\gamma_z}{2}) \delta s_y    
\end{equation}
\begin{equation}
\label{noisez}
\delta\dot{s}_z=2\sqrt{2}\tilde{g} \delta s_y-4\gamma_x \delta s_z     
\end{equation}  
\end{subequations}
Note that, the dynamics for $s_x$ and $s_y$ describe the evolution of the real and imaginary components of the coherence, respectively. It should be emphasized that $ \dot{s}_i$ ($i=\{x,y,z\}$) does not denote a physical velocity in real space; instead, it characterizes the velocity in the Bloch-state space. The dynamics of the population polarization is governed by $ \dot{s}_z$.\\

Sec. \ref{drift} is devoted to the drift and diffusion mechanisms of this two-mode atom–dimer toy model, while Sec. \ref{Frequency Spectrum} analyzes the diagonal elements of the fluctuation spectrum matrix ($\tilde{S}_{ii}$) as functions of the Feshbach detuning, $\epsilon_b$.
\subsection{Drift, and Diffusion Matrix}\label{drift}
 We now obtain the drift matrix \cite{P_many} from Eq. (\ref{noise}),
\begin{equation}
\label{drift matrix}
\Gamma=\begin{pmatrix}
0 & k & 0\\ -k& 0 & -2\sqrt{2}\tilde{g}\\
0 & 2\sqrt{2}\tilde{g} & 0
\end{pmatrix}    
\end{equation}
whose eigen values are
$\lambda_0=0$, and $\lambda_\pm=\pm i\omega$,    
when $\omega=\sqrt{8\tilde{g}^2+k^2}$. 
\begin{figure}
    \centering
    \includegraphics[width=0.8\linewidth]{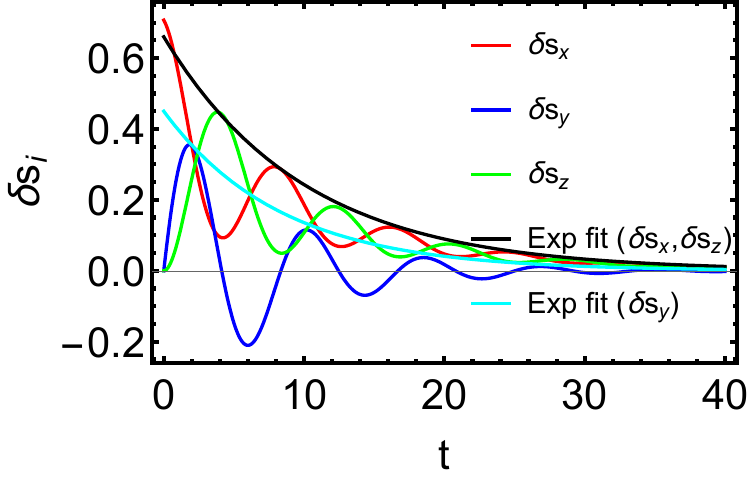}
    \caption[short description]{Real-time dynamics of the Bloch-vector fluctuations $\delta s_x$ (\red), $\delta s_y$ (\blue), and $\delta s_z$ (\green), with exponential fits shown for $\delta s_x$ and $\delta s_z$ (\black), and $\delta s_y$ (\cyan), for finite coupling ($\gamma_x$) and detuning ($\gamma_z$) noise.}
    \label{real dynamics}
\end{figure}

\begin{figure}
\centering
\begin{subfigure}{0.9\linewidth}
    \centering
    \includegraphics[width=\linewidth]{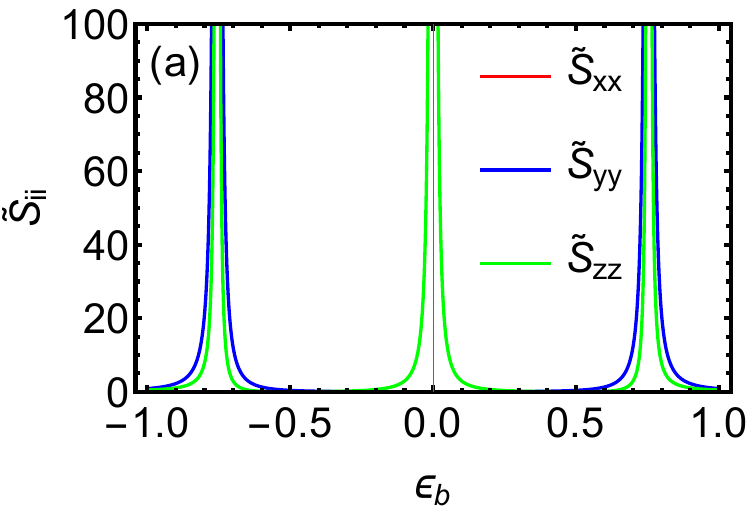}
    \phantomcaption 
    \label{coupling_variance_x}
\end{subfigure}
\begin{subfigure}{0.9\linewidth}
    \centering
    \includegraphics[width=\linewidth]{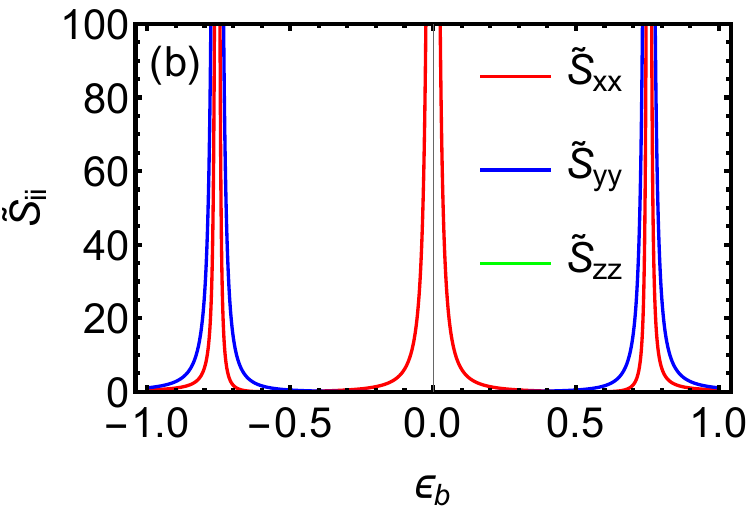}
    \phantomcaption 
    \label{detuning_variance_z}
\end{subfigure}
\caption[short description]{Fluctuation matrix elements ($\tilde{S}_{ii}$) as functions of the Feshbach detuning ($\epsilon_b$). The components $\tilde{S}_{xx}$ (\red), $\tilde{S}_{yy}$ (\blue), and $\tilde{S}_{zz}$ (\green) are indicated by their respective colors. Panels show the cases where noise is present only in (a) the Feshbach coupling ($\gamma_x$) and (b) the detuning ($\gamma_z$).}
\label{variance detuning}
\end{figure}
The diffusion matrix ($D$) \cite{P_many} is now obtained from Eq. (\ref{noise}) \cite{P_many}.
where,\begin{equation}
\label{diagonal diffusion}
D=\begin{pmatrix}
\frac{\gamma_z}{2} & 0 & 0\\
0 & \frac{\gamma_z}{2}+4\gamma_x & 0\\
0 & 0& 4\gamma_x
\end{pmatrix}    
\end{equation} 
In the presence of diffusion, the drift matrix governs the restoring dynamics of the two-state atom-dimer system, driving the fluctuations back toward the steady state after a perturbation.
\subsection{Frequency Spectrum of Fluctuation}\label{Frequency Spectrum}
\begin{subequations}
\begin{equation}
\bigg\langle d \bigg(\delta \mathbf{s}(t)\bigg)d \bigg(\delta \mathbf{s}(0)\bigg)\bigg\rangle
= e^{-\Gamma t}\bigg\langle d \bigg(\delta\mathbf{s}^2(0)\bigg)\bigg\rangle ,
\end{equation}
\begin{equation}
\bigg\langle d \delta \mathbf{s}(0) d \delta \mathbf{s}(t)\bigg\rangle
= \bigg\langle d \bigg(\delta \mathbf{s}^2(0)\bigg)\bigg\rangle e^{-\Gamma^{T} t}.
\end{equation}
\end{subequations}

Here, $\bigg\langle d \bigg(\delta\mathbf{s}(t)\bigg) d\bigg(\delta\mathbf{s}(0)\bigg)\bigg\rangle$ measures how the fluctuation of the Bloch-vector components at a later time remain correlated with their initial ones.

Therefore, by taking the Fourier transform of the autocorrelation function, where $\epsilon_b$ denotes the detuning around the Feshbach resonance point, the low-frequency spectrum can be obtained as \cite{saha2023phase}
\begin{subequations}
\begin{equation}
\tilde{S}(\epsilon_b)
=
\int_{-\infty}^{\infty}
dt e^{i\epsilon_b I t}
\bigg\langle d \bigg(\delta\mathbf{s}(t)\bigg)d \bigg(\delta\mathbf{s}(0)\bigg)\bigg\rangle ,
\end{equation}
\text{Thus, the autocorrelation function of the fluctuations in} {energy-domain becomes,}
\begin{equation}
\tilde{S}(\epsilon_b)
=
\Gamma^{-1}_{\text{left}}
\bigg\langle d\bigg(\delta\mathbf{s}^2(0)\bigg)\bigg\rangle
+
\bigg\langle d\bigg(\delta\mathbf{s}^2(0)\bigg)\bigg\rangle
\Gamma^{-1}_{\text{right}} .
\end{equation}    
\end{subequations}
where, \[\Gamma_{\text{left}(\text{right})}=\Gamma{+(-)} i\epsilon_b I\]
\begin{subequations}
\text{Fluctution matrix $\tilde{S}$ along principle axes become }
\begin{equation}
\tilde{S}=2\Gamma^{-1}_{\text{left}}D\Gamma^{-1}_{\text{right}}    
\end{equation}
\end{subequations}
where, $\tilde{S}_{ii}$ are given by
\begin{subequations}
\begin{equation}
\label{x diffusion}
    \tilde{S}_{xx}=\frac{\gamma_z (\epsilon^2_b-8\tilde{g}^2)^2}{\epsilon^2_b(\epsilon^2_b-\omega^2)^2}
\end{equation}
\begin{equation}
\label{y diffusion}
    \tilde{S}_{yy}=\frac{(\gamma_z+8\gamma_x)\epsilon^2_b}{(\omega^2-\epsilon^2_b)^2}
\end{equation}
\begin{equation}
\label{z diffusion}
\tilde{S}_{zz}=\frac{8\gamma_x(\epsilon^2_b-k^2)^2}{\epsilon^2_b(\omega^2-\epsilon^2_b)^2}    
\end{equation}
\label{all diffusion}
\end{subequations}
 Since Eq. (\ref{all diffusion}) is invariant under $\epsilon_b\to-\epsilon_b$, all spectra are symmetric about $\epsilon_b=0$, yielding equidistant peaks; $\epsilon_b>0$ ($<0$) corresponds to excitation creation (annihilation).

Fig. (\ref{real dynamics}) shows the dynamics of the Bloch-vector fluctuations together with their exponential fits. The oscillatory deviations $\delta s_i$ are bounded by the envelope $\delta s_i(t)=\delta s_i^0 e^{-\Gamma_i t}$,
with $\delta s_x^0=\delta s_z^0=0.66$, $\delta s_y^0=0.45$, $\Gamma_x=\Gamma_z=0.1$, and $\Gamma_y=0.12$.

If only $\gamma_x$ is present in Eq. (\ref{all diffusion}), the noise acts along the $s_x$ axis, leaving $\tilde{S}_{xx}$ unaffected. At $\epsilon_b = 0$, $\gamma_x$ produces a maximal response because the atom-molecule mixing is optimal and is governed solely by the coupling strength. Since $\gamma_x$ directly modifies the polarization velocity, $\dot{s}_z$, $\tilde{S}_{zz}$ exhibits a peak at resonance. As $\epsilon_b \rightarrow 0$, the energy asymmetry vanishes and therefore no longer influences $\dot{s}_y$. For finite $\epsilon_b$, noise-induced transitions compete with the energy bias, giving rise to two off-resonant divergences. Population diffusion is maximized when the detuning matches a nonzero eigenvalue of the drift matrix, leading to the peaks shown in Fig. (\ref{coupling_variance_x}).

If only $\gamma_z$ is present in Eq. (\ref{all diffusion}), the noise acts along the $s_z$ axis, leaving $\tilde{S}_{zz}$ unchanged. Since $\gamma_z$ enters only through the coherence plane, namely through $\dot{s}_x$ and $\dot{s}_y$, the vanishing energy asymmetry at $\epsilon_b = 0$ again leaves $\dot{s}_y$ unaffected. Consequently, only $\tilde{S}_{xx}$ exhibits a resonant peak. For $\epsilon_b \neq 0$, both $\tilde{S}_{xx}$ and $\tilde{S}_{yy}$ develop two off-resonant peaks corresponding to the nonzero eigenvalues of the drift matrix, as shown in Fig. (\ref{detuning_variance_z}).

The initial conditions, and parametrs for the numerical solutions are discussed in Appendix \ref{initial state}.\\

In Sec. \ref{auto correlation time}, we analyze the velocity response from Force-damped harmonic oscillator model.
\section{Auto-correlation function of the Bloch vector components}\label{auto correlation time}
Sec. \ref{both type mobility} examines the resistive and reactive parts of the dynamic mobility, whereas Sec. \ref{static mobility section} determines the temperature via the Einstein relation of static mobility.
\subsection{Dynamic Mobility}\label{both type mobility}
\begin{figure}
    \centering
    \includegraphics[width=0.8\linewidth]{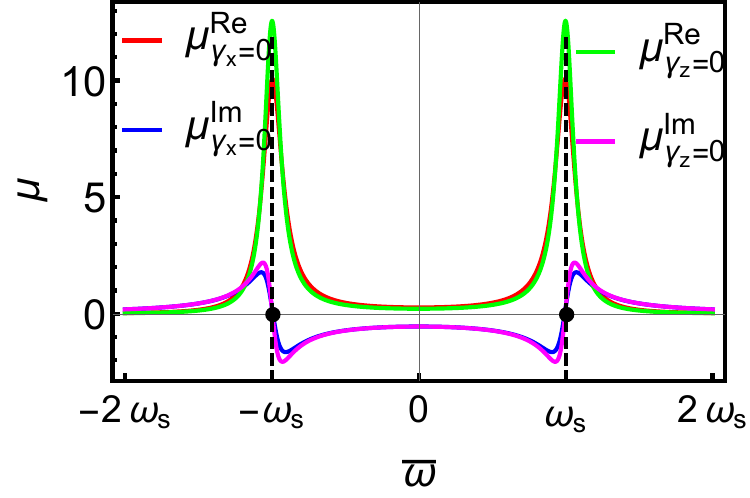}
    \caption[short description]{Kubo spectral mobility, $\mu(\bar{\omega})$, versus driving frequency $\bar{\omega}$. The real part, $\mu^{\mathrm{Re}}$, is shown for ($\gamma_x \neq 0$, $\gamma_z = 0$) (\green) and ($\gamma_z \neq 0$, $\gamma_x = 0)$ (\red), while the imaginary part, $\mu^{\mathrm{Im}}$, is shown for ($\gamma_x \neq 0$, $\gamma_z = 0$) (\magenta) and $(\gamma_z \neq 0$, $\gamma_x = 0$) (\blue). Here, $\gamma_x$ and $\gamma_z$ represent noise in the Feshbach coupling $\tilde{g}$ and the detuning $\epsilon_b$, respectively. The symbols (\blackbullet) mark $\mu^{\mathrm{Im}}$ at the dynamic resonance frequencies $\pm\omega_s$.}
    \label{mobility}
\end{figure}
\begin{figure}
    \centering
    \includegraphics[width=0.8\linewidth]{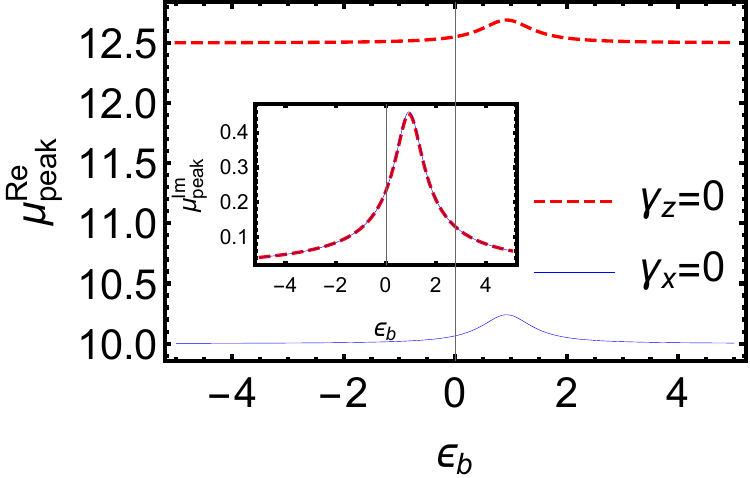}
    \caption[short description]{Dependence of the dynamic resonance peak $\mu^{\mathrm{Re}}_{\mathrm{peak}}$ on the Feshbach detuning $\epsilon_b$. The (\reddashed) and (\blue) curves correspond to coupling noise ($\gamma_x \neq 0$, $\gamma_z=0$) and detuning noise ($\gamma_z \neq 0$, $\gamma_x=0$), respectively. The inset displays the corresponding imaginary component, $\mu^{\mathrm{Im}}_{\mathrm{peak}}$.}
    \label{peak}
\end{figure}
We now obtain the closed-form equations governing the linearized fluctuations of the Bloch components from Eq. (\ref{noise}).
\begin{subequations}
\begin{equation}
\label{actual}
\delta\dddot {s}_i+\gamma \delta \ddot{s}_i+\tilde\omega^2 \delta \dot{s}_i+\tilde{c} \delta s_i=0, \quad \text{where} \quad i\in \{x,y,z\}   
\end{equation}
\text{if $ \delta \dot{s}_i=\delta v_i$, then we obtain}
\begin{equation}
\label{velocity1}
\delta \ddot{v}_i+\gamma \delta \dot{v}_i+\tilde\omega^2 \delta v_i=-\tilde{c}\int \delta v_i dt 
\end{equation}
Note that the natural frequency is $\tilde{\omega}^{2}=\omega^{2}+\gamma^{2}/4$, with damping coefficient $\gamma=8\gamma_x$ ($\gamma_z$) for coupling (detuning) noise, and constant force $\tilde{c}$.
\begin{equation}
        \int \delta s_j dt= \text{constant}, \quad \text{and}\quad \int \delta s_y dt= 0
    \end{equation}
    \text{where $j\in\{x,z\}$. So, Eq. (\ref{velocity1}) becomes}
    \begin{equation}
    \label{modified velocity}
   \delta\ddot{v}_i+\gamma \delta\dot{v}_i+\tilde\omega^2 \delta v_i=\text{constant}     
    \end{equation}
\end{subequations}
  Introducing the equilibrium velocity autocorrelation function,
$\mu_i(t)=\langle \delta v_i(0)\delta v_i(t)\rangle_{\mathrm{eq}}$,
and employing Eq. (\ref{modified velocity}), one obtains
\\
\begin{subequations}
   \begin{equation}
    \label{new velocity}
   \ddot{\mu}_i(t)+\gamma \dot{\mu}_i(t)+\tilde\omega^2 \mu_i(t)=\text{constant}     \end{equation}
\text{Solving Eq. (\ref{new velocity}), we obtain, as shown in} \cite{P_many},
\begin{equation}
    \mu_i(t)=c^\prime_1e^{-\lambda_+ t}+c^\prime_2e^{-\lambda_- t}
\end{equation}
\text{where time independent coefficients are},
{$c^\prime_1=(\lambda_- \mu_i(0)+\dot\mu_i(0))/(\lambda_--\lambda_+)$,} { $ c^\prime_2=(\lambda_+ \mu_i(0)+ \dot{\mu}_i(0))/(\lambda_+-\lambda_-)$, and simple poles at $\lambda_\pm=\gamma/2\pm i\omega_s$.} {Using the value of these, we obtain}
 \begin{equation}
 \label{velocity}
\mu_i(t)=e^{-\frac{\gamma  t}{2}}\sin(\omega_st+\theta_i)     
 \end{equation} 
\end{subequations}

 where, phase ($\theta_i$) is the angle between driving force and resultant motion,
 $\sin\theta_i=1$,  $\cos\theta_i=\gamma/2\omega_s$, and the quantity
$\omega_s=\sqrt{\tilde{\omega}^2-\gamma^2/4}$
denotes the damped angular frequency. It governs the oscillatory part of the dynamics and determines the positions of the zero crossings (null points) in the damped oscillatory system \cite{P-171}.

Now, the normalized mobility in time-domain is defined as
\begin{figure}
    \centering
    \includegraphics[width=0.8\linewidth]{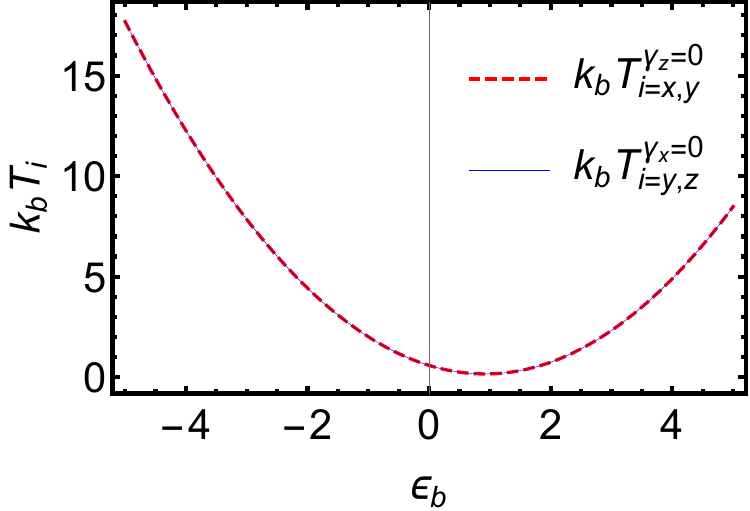}
    \caption[short description]{Temperature ($T_i$) as a function of detuning ($\epsilon_b$) for coupling noise only ($\gamma_x \neq 0$, $\gamma_z = 0$; \reddashed) and detuning noise only ($\gamma_z \neq 0$, $\gamma_x = 0$; \blue).}
    \label{temp}
\end{figure}
 \begin{subequations}
  \begin{equation}
    \label{unitless}
 \mu_i(t)=e^{-\frac{\gamma t}{2}}\sin(\omega_s t+\beta_i)    
    \end{equation}
{where, $\mu_i(0)=\sin\theta$. AC mobility, $\mu(\bar\omega)$ is given by}
\begin{equation}
\mu(\bar\omega)=\int^\infty_0 \mu_i(t) e^{i\bar\omega t} dt    
\end{equation}
{Under an external sinusoidal force, $F_0 e^{i\bar{\omega} t}$, oscillating at frequency $\bar{\omega}$ \cite{P65}, the mobility is defined as $\mu(\bar{\omega})=v(\bar{\omega})/F(\bar{\omega})$, implying that the response is measured at the driving frequency \cite{p59}. Now, the response is retarded, reflecting the causal nature of the dynamics \cite{P563}.
}
 {Its gain (real) and loss (imaginary) parts can be defined as},
\begin{equation}
\label{re}
 \mu^{\text{Re}}(\bar\omega)=\frac{\gamma}{2}\bigg(\frac{2+\frac{\bar\omega}{\omega_s}}{(\frac{\gamma}{2})^2+(\bar\omega+\omega_s)^2}+\frac{2-\frac{\bar\omega}{\omega_s}}{(\frac{\gamma}{2})^2+(\bar\omega-\omega_s)^2}\bigg)   \end{equation}
 \begin{equation}
 \label{Im}
 \mu^{\text{Im}}(\bar\omega)=\frac{\gamma}{2}\bigg(\frac{\frac{\gamma}{4\omega_s}-\frac{\bar\omega+\omega_s}{\gamma}}{(\frac{\gamma}{2})^2+(\bar\omega+\omega_s)^2}+\frac{\frac{\gamma}{4\omega_s}+\frac{\bar\omega-\omega_s}{\gamma}}{(\frac{\gamma}{2})^2+(\bar\omega-\omega_s)^2}\bigg)\end{equation}
 \label{mu all}
 \end{subequations}
\subsubsection{Resonant Enhancement of the Response Function Near the Dynamic Resonance}
The real and imaginary parts of the mobility, $\mu^{\mathrm{Re}}(\bar{\omega})$ and $\mu^{\mathrm{Im}}(\bar{\omega})$, form a Hilbert-transform pair and are related through the Kramers-Kronig relations.  The real part, $\mu^{\mathrm{Re}}(\bar{\omega})$, characterizes the reactive (dispersive) response and corresponds to the component oscillating in phase with the external drive, whereas the imaginary part, $\mu^{\mathrm{Im}}(\bar{\omega})$, describes the absorptive (dissipative) response and represents the out-of-phase component. Since $\mu^{\mathrm{Re}}(\bar{\omega})$ is associated with reversible response, analogous to the elastic response of a damped spring-mass oscillator, it generally exhibits a larger amplitude than $\mu^{\mathrm{Im}}(\bar{\omega})$. In contrast, $\mu^{\mathrm{Im}}(\bar{\omega})$ originates from the phase lag between the response and the driving force and therefore quantifies dissipation, dephasing, and irreversible energy loss, as shown in Fig. (\ref{mobility}).
 At $\bar{\omega}=\omega_s$, the system reaches dynamic resonance, where the power absorbed from the external driving force exactly balances the power dissipated through damping \cite{P14} for which $\mu^{\text{Im}}(\bar\omega)=0$. For a driving force of amplitude $F_0$, both $\mu^{\mathrm{Re}}(\bar{\omega})$ (in-phase component) and $\mu^{\mathrm{Im}}(\bar{\omega})$ (out-of-phase component) exhibit two distinct peaks associated with the two eigenmodes, corresponding to the poles of Eqs. (\ref{re}) and (\ref{Im}). The valley separating these peaks corresponds to a frequency-mismatch region, where the oscillation is substantially out of phase with the driving field, resulting in a suppressed response \cite{Taylor2005}. The sign change of $\mu^{\mathrm{Im}}(\bar{\omega})$ signifies a reversal of the phase difference, $\Phi(\bar{\omega})$ between $F(\bar{\omega})$ and the induced velocity, ($v(\bar\omega)$).\\
 The initial condition for finite coupling and that for detuning noise are listed in Appendices \ref{non zero coupling} and \ref{non zero detuning}, respectively.
 \subsubsection{Emergence of a Pronounced Response-Function Peak Near the Feshbach Resonance}
 Near a Feshbach resonance, the magnetically tunable scattering length is given by \cite{chin2010feshbach}
\begin{equation}
    a_s(B)=a_{\text{bg}}\left(1-\frac{\Delta B}{B-B_0}\right),
    \label{scattering}
\end{equation}

where $a_{\text{bg}}$, $\Delta B$, and $B_0$ denote the background scattering length, resonance width, and resonance position, respectively.

 In Eq. (\ref{scattering}), $a_s(B)$ diverges near the Feshbach resonance, rendering the system extremely sensitive to perturbations. Therefore, $\mu^{\text{Re(Im)}}_{\text{peak}}$ reaches its maximum in the vicinity of the Feshbach resonance, as shown in Fig. (\ref{peak}).
\subsection{Extracting Temperature from Static Mobility}\label{static mobility section}
Static mobility, $\mu_0$, is defined as \cite{P_many}
\begin{subequations}
\begin{equation}
\label{mobility static}
\mu_0=\int_0^\infty \mu_i(t)dt=\frac{\gamma}{\tilde{\omega}^2}.
\end{equation}
\text{We define an effective temperature as \cite{p323}}
  \begin{equation}
\label{Einstein relation}
k_B T_i=\frac{D_{ii}}{\mu_0},\quad \text{where} \quad i\in\{x,y,z\}.
\end{equation}  
\end{subequations}

This is not the physical temperature. Substituting the expression for $\mu_0$ from Eq. (\ref{mobility static}) into Eq. (\ref{Einstein relation}), we obtain the effective temperatures. The temperatures corresponding to the different noise strengths are
\begin{subequations}
   \begin{equation}
k_B T_j=
\begin{cases}
\dfrac{\tilde{\omega}^2}{2}, & j\in\{x,y\}, \\
0, & j=z,
\end{cases}
\qquad (\gamma_x=0),
\end{equation}
and
\begin{equation}
k_B T_j=
\begin{cases}
\dfrac{\tilde{\omega}^2}{2}, & j\in\{y,z\}, \\
0, & j=x,
\end{cases}
\qquad (\gamma_z=0).
\end{equation} 
\end{subequations}
\subsubsection{Effective Temperature: A Physical Perspective}
A physical temperature can be defined in equilibrium systems \cite{P65}. An effective temperature can also be introduced, providing an equilibrium-like description of the system \cite{vallejo2021qubit}. Analogously, we introduce an effective temperature $T$, defined through the ratio of fluctuations to mobility, despite the system being out of equilibrium. The directional dependence obtained here reflects the anisotropy of the noisy environment. 

  Large restoring forces correspond to higher oscillation frequencies of the coherence and imbalance velocity fluctuations. Such large (small) restoring forces occur far from (near) the Feshbach resonance, i.e., when $\epsilon_b$ is large (small). Near the Feshbach resonance ($\epsilon_b \approx 0$), the populations of the atomic and molecular states become nearly equal because the energy cost associated with converting one species into the other is minimal. Consequently, the strong hybridization between these two states reduces the restoring force required to return the system to its equilibrium configuration. As a result, the oscillation frequencies associated with the coherence and imbalance velocity fluctuations attain their minimum values. As $\epsilon_b$ increases, this near-degeneracy is lifted, weakening the atom-molecule hybridization and increasing the restoring force. Consequently, the oscillation frequencies increase with increasing $\epsilon_b$.

The minimum in $T$ can also be understood in terms of mobility. Near $\epsilon_b \approx 0$, the susceptibility of the two-mode system to periodic driving is maximal, leading to a peak in the DC mobility $\mu_0$ [Eq. (\ref{Einstein relation})]. Physically, moving away from $\epsilon_b\approx 0$ reduces the system's susceptibility and transport efficiency.

\subsubsection{Noise-Induced Modifications of the Effective Temperature}

For $\gamma_z \neq 0$ and $\gamma_x=0$, the noise couples only to the coherence dynamics, $\dot{s}_x$ and $\dot{s}_y$, and leaves the imbalance dynamics, $\dot{s}_z$, unaffected. Consequently, the fluctuations remain confined to the $s_x$ - $s_y$ plane, yielding equal effective temperatures along the transverse directions, $T_x=T_y$, while no effective temperature is generated along the longitudinal direction, $T_z=0$, as shown in Fig. (\ref{temp}).

In contrast, $\gamma_x$ noise drives the out-of-phase coherence and polarization dynamics, $\dot{s}_y$ and $\dot{s}_z$, while leaving $\dot{s}_x$ unaffected. As a result, $T_y=T_z$, whereas no effective temperature is induced along the $s_x$ direction, yielding $T_x=0$ [Fig. \ref{temp}].

In Sec. \ref{feature}, we discuss several characteristic features of the force-damped oscillator. In this section, we investigate their behavior near the Feshbach resonance.
\section{Few Features of Force-damped Harmonic Oscillator}\label{feature}

Sec. \ref{phase} discusses how the system synchronizes with the external drive by analyzing the phase lag between the response and the drive. Sec. \ref{power section} examines the exchange of power between the system and the periodic drive. The quality factor and the resonance width of the dynamical resonance, in relation to the Feshbach resonance, are discussed in Secs. \ref{quality factor section} and \ref{Full Width Half Maxima section}, respectively.
\begin{figure}
    \centering
    \includegraphics[width=0.8\linewidth]{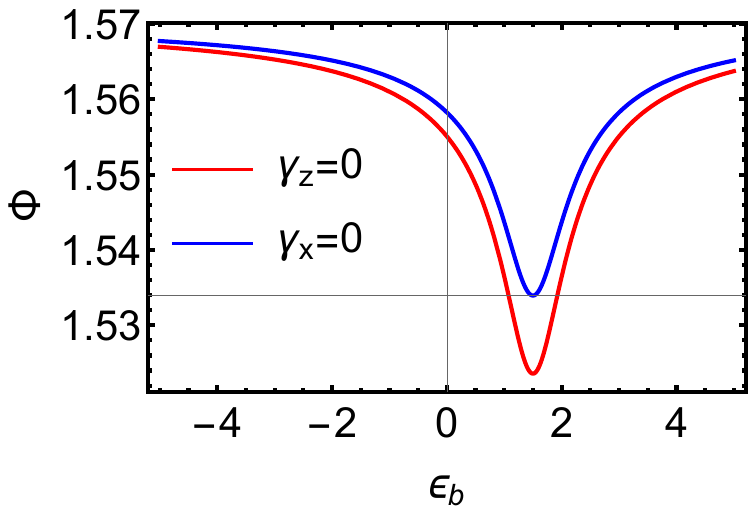}
    \caption[short description]{Phase difference ($\Phi$) plotted as a function of detuning ($\epsilon_b$) for coupling noise ($ \gamma_x\neq0,\gamma_z=0$) and detuning noise ($\gamma_z\neq0,\gamma_x=0$), denoted by the (\red) and (\blue) curves, respectively.}
    \label{phase detuning}
\end{figure}
\subsection{Phase Synchronization Across the Feshbach Resonance}\label{phase}

Since $\mu^{\text{Re}}$ and $\mu^{\text{Im}}$ correspond to the in-phase and out-of-phase components of the response to the periodic drive, respectively, the phase difference $\Phi(\bar{\omega})$ characterizes the phase lag between the system response and the external drive. It is given by \cite{P76}
\begin{equation}
\Phi(\bar{\omega})=
\tan^{-1}\left(
\frac{\mu^{\text{Im}}(\bar{\omega})}
     {\mu^{\text{Re}}(\bar{\omega})}
\right).
\label{phase definition}
\end{equation}
A smaller phase difference indicates stronger synchronization, implying a closer balance between the driving force and the spring force \cite{morin2008classical}. In Fig. (\ref{phase detuning}) near $\epsilon_b \approx 0$, atom-dimer conversion is maximized due to the near degeneracy of the atomic and molecular states and the strongest coupling between the two channels. As a result, the phase difference defined in Eq. (\ref{phase definition}) attains a minimum, irrespective of whether the noise affects the coupling or the detuning.
\subsection{Power Evolution in the Vicinity of a Feshbach Resonance}\label{power section}
When ($\bar{\omega}=\omega_s$), the frequency of the external periodic force matches the intrinsic frequency of the system, leading to dynamic resonance. As a result, the reversible response, $\mu^{\mathrm{Re}}(\bar{\omega})$, is maximized, whereas the irreversible response, $\mu^{\mathrm{Im}}(\bar{\omega})$, is minimized near $\bar{\omega}=\omega_s$ \cite{Taylor2005}. 
\begin{figure}
    \centering
    \includegraphics[width=0.8\linewidth]{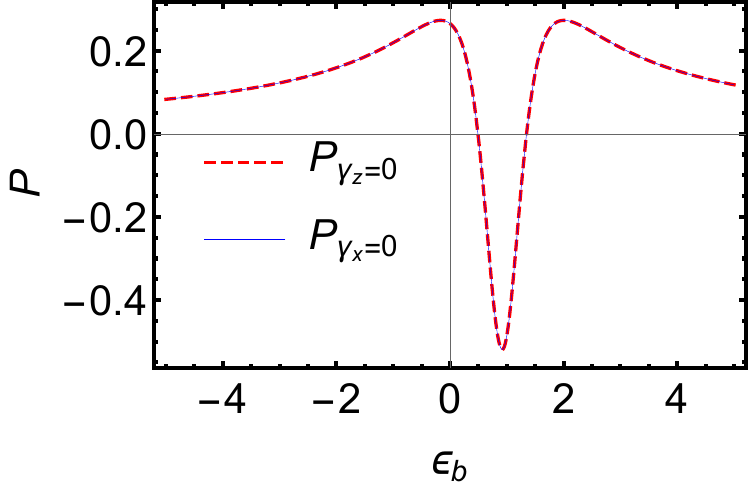}
    \caption[short description]{Power ($P$) versus detuning ($\epsilon_b$) at dynamic resonance point ($\bar{\omega}=\omega_s$) for the cases of exclusive coupling noise ($\gamma_x \neq 0$, $\gamma_z = 0$) and exclusive detuning noise ($\gamma_z \neq 0$, $\gamma_x = 0$). The corresponding curves are shown as (\reddashed) and (\blue) lines, respectively.}
    \label{power}
\end{figure}
The cycle-averaged absorbed power is given by
\begin{equation}
\langle P\rangle=\frac{|F_0|^2}{2}\mu^{\mathrm{Im}}(\bar\omega)
\end{equation}.
$\langle P\rangle>0$ ($<0$) corresponds to absorption from (deliver to) the driving field, i.e., $\mu^{\mathrm{Im}}(\bar{\omega})>0$ ($<0$). Since $\mu^{\mathrm{Im}}(\bar{\omega})$ is asymmetrically distributed about the positive and negative sides of the vertical axis, it indicates irreversible energy dissipation in the system. As shown in Fig. (\ref{power}), $\lvert \langle P \rangle \rvert$ is maximized near $\epsilon_b=0$, where the system is most sensitive to noise and $\mu^{\text{Im}}(\bar{\omega})$ attains its peak value.

\begin{figure}
    \centering
    \includegraphics[width=0.8\linewidth]{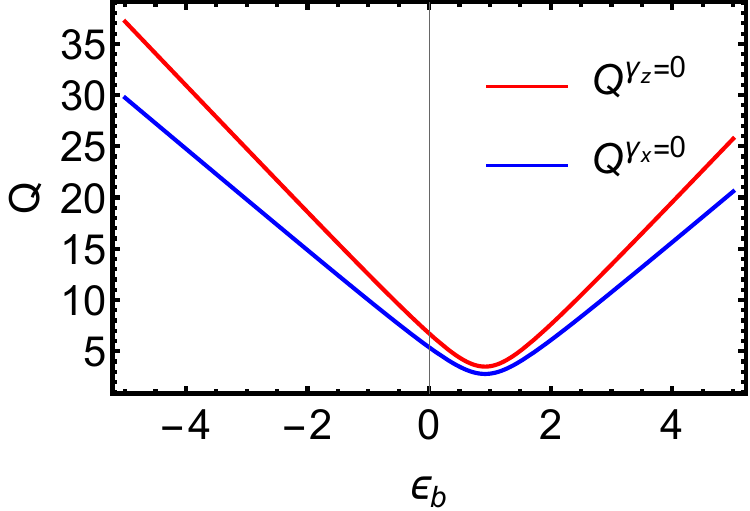}
    \caption[short description]{Quality factor ($Q$) plotted against detuning ($\epsilon_b$) for exclusive coupling noise ($\gamma_x \neq 0$, $\gamma_z= 0$; \red) and exclusive detuning noise ( $\gamma_z \neq 0$, $\gamma_x = 0$; \blue).}
    \label{quality}
\end{figure}
\subsection{Quality-factor Characteristics Around the Feshbach Resonance}\label{quality factor section}
The quality factor ($Q$), a measure of dynamic resonance sharpness \cite{P261}, equals the energy stored per cycle divided by the energy dissipated per cycle \cite{P129}. Hence, a larger quality factor $Q$ indicates weaker dissipation and a more weakly damped oscillation \cite{P304}. As a result, the oscillation amplitude decreases by a factor of $e^{-\pi/Q}$ per cycle \cite{P264}. It is defined as
\begin{equation}
Q=\frac{\omega_s}{\gamma}.
\end{equation}
As shown in Fig. (\ref{quality}), the quality factor $Q$ attains its minimum value at the Feshbach resonance ($\epsilon_b = 0$), where the diverging scattering length enhances noise-induced fluctuations, leading to maximal energy dissipation \cite{P-527}. At $\epsilon_b = 0$, the dimer is weakly bound and possesses a large spatial extent, rendering the system particularly susceptible to fluctuations. Consequently, fluctuation-induced effects are most pronounced in this regime, resulting in the maximal suppression of $Q$. As the system is detuned away from the Feshbach resonance, the dissipation decreases and $Q$ correspondingly increases. 

\subsection{Full Width Half Maxima Across the Feshbach Resonance}\label{Full Width Half Maxima section}
If $\bar\omega=\pm\omega_s$ then both $\mu^{\text{Re (Im)}}(\pm\omega_s)$ obtain its peak value, 
\begin{subequations}
 \begin{equation}
  \mu^{\text{Re}}_{\text{peak}}(\pm\omega_s)=\frac{\gamma}{2}\bigg(\frac{3}{(\frac{\gamma}{2})^2+(2\omega_s)^2}+\frac{1}{(\frac{\gamma}{2})^2}\bigg)  
\end{equation}
\begin{equation}
 \mu^{\text{Im}}_{\text{peak}}(\pm\omega_s)=\frac{1}{2\omega_s}\bigg(1+\frac{(\frac{\gamma}{2})^2-2\omega^2_s}{(\frac{\gamma}{2})^2+(2\omega_s)^2}\bigg)  
\end{equation} 
\begin{equation}
  \mu^{\text{Re (Im)}}(\bar\omega)=\frac{\chi^{\text{Re  (Im)}}_{\text{peak}}(\omega_s)}{2}   
\end{equation}
\end{subequations}
The dynamic resonance width can be extracted by solving the above equation for the two half-maximum frequencies, $\bar{\omega}_+$ and $\bar{\omega}_-$. FWHM is subsequently obtained as
$\mathrm{FWHM}=\bar{\omega}_+ - \bar{\omega}_-$ \cite{P14}. At $\bar{\omega}_{\pm}$, the amplitude falls to approximately 70\% of its peak value \cite{P34}. 

\begin{figure}
\centering
\begin{subfigure}{0.9\linewidth}
    \centering
    \includegraphics[width=\linewidth]{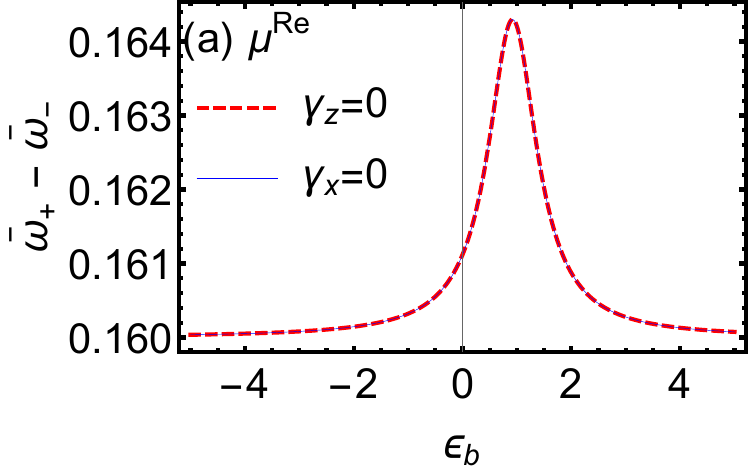}
    \phantomcaption 
    \label{mu_re}
\end{subfigure}
\begin{subfigure}{0.9\linewidth}
    \centering
    \includegraphics[width=\linewidth]{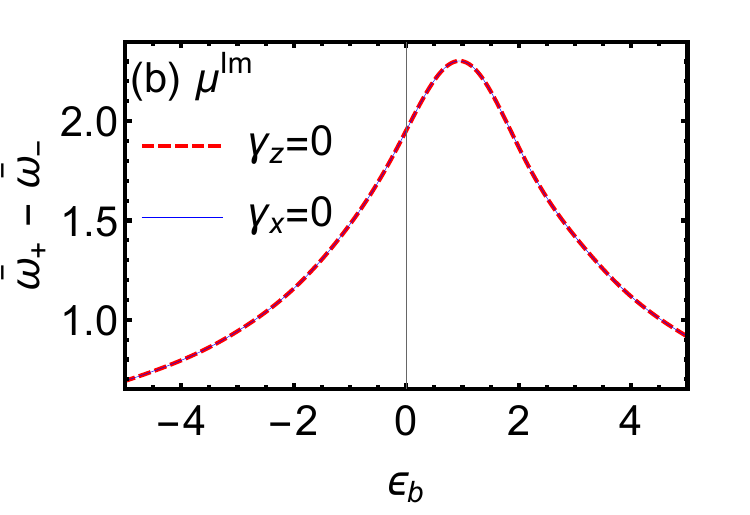}
    \phantomcaption 
    \label{mu_im}
\end{subfigure}
\caption[short description]{Dynamic resonance linewidth, $(\bar{\omega}_{+}-\bar{\omega}_{-})$, as a function of the Feshbach detuning $\epsilon_b$. The linewidths extracted from the response function are shown for (a) the real part, $\mu^{\mathrm{Re}}(\bar{\omega})$, and (b) the imaginary part, $\mu^{\mathrm{Im}}(\bar{\omega})$. The (\reddashed) curves correspond to coupling noise ($\gamma_x \neq 0$, $\gamma_z = 0$), while the (\blue) curves correspond to detuning noise ($\gamma_z \neq 0$, $\gamma_x = 0$). 
}
\label{mu mobility}
\end{figure}

Near the Feshbach resonance ($\epsilon_b \approx 0$), the atom-dimer interconversion rate becomes maximal due to the strong coupling between the open and closed channels \cite{Inguscio2007}. Consequently, the lifetime of particles in either the atomic or dimeric state is reduced. Furthermore, near a Feshbach resonance, the scattering length diverges, resulting in a significantly enhanced collision rate owing to the large elastic scattering cross-section \cite{Inguscio2007}. These effects increase the damping rate and shorten the lifetime of the excitation, resulting in a broader dynamic resonance. Therefore, the FWHM attains its maximum value near $\epsilon_b = 0$, as shown in Fig. (\ref{mu mobility}). Since energy is absorbed over a broader frequency range in $\mu^{\text{Im}}(\bar\omega)$ (Fig. \ref{mobility}), the FWHM of $\mu^{\text{Im}}(\bar\omega)$ is larger than that of $\mu^{\text{Re}}(\bar\omega)$. Since $\mu^{\text{Re}}(\bar\omega)$ vanishes more rapidly than $\mu^{\text{Im}}(\bar\omega)$ away from dynamic resonance [Eqs. (\ref{mu_re})
and (\ref{mu_im})], the dispersive response is more localized around the dynamic
resonance frequency, $\omega_s$. Therefore, the FWHM of
$\mu^{\text{Re}}(\bar\omega)$ is smaller than that of the dissipative. The broader dynamic resonance peak in the imaginary part reflects weaker temporal correlations (shorter memory) between the initial and final states. As a result, the peak mobility is reduced, and the response is spread over a broader frequency range.

Notably, Figs. (\ref{mobility}) - (\ref{mu mobility}) indicate that coupling noise and detuning noise have qualitatively similar effects. Specifically, both noise sources lead to the same underlying dynamical behavior in the vicinity of the dynamic resonance as well as the Feshbach resonance.

As discussed in \cite{quantumgas2}, the slight rightward shift of the Feshbach resonance position away from $\epsilon_b=0$, shown in Figs. (\ref{temp}) - (\ref{mu mobility}), is attributed to magnetic-field fluctuations in the Feshbach detuning and thermal effects in the Feshbach coupling. 
\section{summary}\label{conclusion}
We consider a bosonic atom-molecule system in which a pair of bosonic atoms can associate to form a bosonic molecule through a Feshbach resonance. The Feshbach detuning quantifies the energy difference between the free atomic pair and the molecular bound state. In this model, both the Feshbach coupling and the detuning are subject to Gaussian white noise, allowing us to investigate the influence of stochastic fluctuations on the dynamics of the atom-molecule condensate.

Experimentally, coherent atom-molecule superpositions can be realized using magnetic-field pulses near a Feshbach resonance, with the relative phase extracted from absorption-imaging measurements of the final atomic and molecular populations \cite{donley2002atom}. In the context of our theoretical description, such coherent conversion is represented by the Feshbach coupling term. Fluctuations in the coupling strength, arising from thermal or technical noise sources, can manifest as relative-phase fluctuations between weakly coupled condensates \cite{gati2006noise}. Likewise, magnetic-field fluctuations directly affect the resonance condition and therefore act as detuning noise. Controlled modulation of the detuning has been employed experimentally to induce photon-assisted atom-molecule conversion even above the molecular dissociation threshold \cite{thompson2005ultracold}.

 We find that coupling and detuning noise generate distinct anisotropic fluctuation spectra. At Feshbach resonance, coupling noise drives population diffusion, producing fluctuation peaks along the imbalance axis, whereas detuning noise induces phase diffusion, enhancing fluctuations along the symmetric coherence axis. Away from resonance, coupling noise competes with the bias, giving rise to two off-resonant diffusion maxima and amplified fluctuations in the asymmetric coherence axes and imbalance, while detuning noise continues to produce pure phase diffusion, enhancing only the symmetric and antisymmetric coherence axes. These anisotropic spectral features provide direct signatures of thermal phase fluctuations and decoherence. Experimentally, such fluctuations can be accessed through repeated measurements of relative phase and population imbalance, analogous to matter-wave interferometric measurements of phase fluctuations in split condensates \cite{Schumm2005} and number-fluctuation measurements via absorption imaging \cite{Esteve2008}. In a Feshbach-coupled atom--molecule condensate, the corresponding coherence and imbalance fluctuation spectra can therefore be directly reconstructed from shot-to-shot phase and population measurements.

In our framework, finite-time correlations of coherence and imbalance-velocity fluctuations follow damped-harmonic-oscillator dynamics, with a restoring energy that scales with the effective temperature.
 Under periodic driving, their response is characterized by the mobility, whose real and imaginary parts describe reversible and irreversible dynamics, respectively. Dynamical resonance occurs when the intrinsic and driving frequencies coincide. Phase synchronization is signaled by a minimum phase lag, corresponding to maximal atom–dimer hybridization and power absorption. Consequently, the linewidth is maximal and the quality factor minimal at Feshbach resonance, while weaker hybridization away from resonance narrows the linewidth and increases the quality factor.  Thus, we demonstrate that the Feshbach resonance can control both the fluctuation dynamics and the response of a noisy quantum system.

Experimentally, \cite{stenger1999bragg} used two-photon stimulated Bragg scattering as a spectroscopic probe of the linewidth of a $^{23}\mathrm{Na}$ condensate. Similarly, in the our system, a periodic drive may be employed to study its influence on the fluctuation spectra of the coherence and atom--molecule imbalance velocities.

 \section{Acknowledgements}
AM would like to acknowledge University
Grants Commission (UGC), Govt. of India for financial
support (Student ID: 201610064840). RD would like to acknowledge Science and Engineering Research Board (SERB), currently, Anusandhan
National Research Foundation (ANRF), Department of Science and Technology, Govt. of India for
providing support under the CRG scheme (CRG/2022/007312), and Rashtriya Uchchatar Shiksha Abhiyan (RUSA) 2.0 (Ministry of Education, Govt. of India).
\setcounter{equation}{0}
\appendix
\section{Construction of Mean-Field Dynamics for Bloch Vector Components}\label{appendix MF}
We arrive at the Bloch equations similar to the well-known Nuclear Magnetic Resonance (NMR) process \cite{bloch1946nuclear,viola2000stochastic}
\begin{subequations}
    \begin{equation}
    \label{final1}
    \begin{split}
       \dot{s_x}=2c_1s_ys_z+c_2s_y
        \end{split}
    \end{equation}
    \begin{equation}
    \label{final2}
    \begin{split}
        \dot{s_y}=&-2c_1s_xs_z-c_2s_x-\frac{\tilde{g}}{\sqrt{2}}(1+2s_z-3s^2_z)
        \end{split}
    \end{equation}
    \begin{equation}
    \label{final3}
    \begin{split}
      \dot{s_z}=2\sqrt{2}\tilde{g}s_y
       \end{split}
    \end{equation}
    \label{final}
\end{subequations}
where  $c_1 = U_3/2\hbar - U_1/2\hbar - U_2/8\hbar$ and $c_2 = U_1/\hbar - U_2/4\hbar - \epsilon_b/\hbar$, with both coefficients expressed in terms of the experimental parameters $u_1$, $u_2$, $u_3$, and $\epsilon_b$.

From Eq. (\ref{final}), we obtain two equilibrium points, $(0,0,1)$ and $(0,0,-1/3)$. Among these, the second equilibrium point is more stable, as discussed in \cite{quantumgas2}. Since $s_i = s_i^{\mathrm{eq}} + \delta s_i$, where $s_i^{\mathrm{eq}}$ denotes the equilibrium value of $s_i$, while $\delta s_i$ represents the fluctuation about the equilibrium state. Linearizing Eq. (\ref{final}) around the stable equilibrium point $(0,0,-1/3)$, and neglecting constant terms, we obtain the linearized system:
\begin{subequations}
\label{linearization}
 \begin{equation}
 \delta\dot{s}_x=k \delta s_y   
 \end{equation}
 \begin{equation}
\delta\dot{s}_y=-k \delta s_x-2\sqrt{2}\tilde{g} \delta s_z
 \end{equation}
 \begin{equation}
     \delta\dot{s}_z=2\sqrt{2}\tilde{g} \delta s_y
 \end{equation}
\end{subequations}
Here,
$k=(4U_1/3\hbar-U_2/6\hbar-U_3/3\hbar-\epsilon_b/\hbar)$,
while $c_1$ and $c_2$ are functions of $U_i$, with $i\in\{x,y,z\}$.

If $\tilde{g}$, and $\epsilon_b$ corrupted by Gaussian white noise, $\eta_x$, and $\eta_z$, then we obtain\\
\begin{subequations}
\label{wiener}
\begin{equation}
 d(\delta s_x)-k \delta s_y dt=\delta s_ydw_z   
\end{equation}
 \begin{equation}
d (\delta s_y)+(k \delta s_x+2\sqrt{2}\tilde{g} \delta s_z)dt=-\delta s_xdw_z-2\sqrt{2} \delta s_zdw_x     
 \end{equation}
\begin{equation}
 d (\delta s_z)=2\sqrt{2}\tilde{g} \delta s_y dt+2\sqrt{2} \delta s_ydw_x     
\end{equation} 
\end{subequations}
where, $\eta_i=dw_i/dt$, and $w_i$ is the wiener process when $i\in\{x,z\}$. Generalizing from Eq. (\ref{wiener}), we obtain \cite{P_many}
\begin{equation}
\label{differential format}
d(\delta \mathbf{s}) + \Gamma \delta \mathbf{s} dt =\delta\mathbf{s} d\mathbf{w} ,
\end{equation}
where $\Gamma$ is the drift matrix.
\section{Initial Configuration and Physical Parameters}\label{initial state}

The resulting Bloch equations closely resemble those found in NMR systems \cite{bloch1946nuclear,viola2000stochastic}. The coherent dynamics of the Bloch vector components follow the Heisenberg equation of motion \cite{P189}.
Initial condition for the Bloch vector components can be obtained in the following semiclassical form:
        $\hat{L}_x={(1-z)\sqrt{1+z}}\cos{\tilde\phi}/\sqrt{2}$, 
        $\hat{L}_y={(1-{z})\sqrt{1+z}}\sin{\tilde\phi}/{\sqrt{2}}$, and $
        \hat{L}_z=z$. In this formulation, the atomic and molecular populations, $N_a$ and $N_b$, correspond to the expectation values of the number operators $\hat{a}^\dagger \hat{a}$ and $\hat{b}^\dagger \hat{b}$, respectively, with $\hat{a}=\sqrt{N_a}\,e^{i\tilde\theta_a}$ and $\hat{b}=\sqrt{N_b}\,e^{i\tilde\theta_b}$. The population polarization and relative phase are
$z=2\hat{b}^{\dagger}\hat{b}-\hat{a}^{\dagger}\hat{a}$ and
$\tilde\phi=2\tilde\theta_a-\tilde\theta_b$, respectively. In the non-rigid pendulum picture, the phase-space point $(z,\tilde\phi)=(0,0)$ corresponds to the Josephson $\mathbf{0}$-state \cite{marino1999bose,saha2023phase}. Here, $z$ and $\tilde\phi$ constitute a canonically conjugate pair in the classical phase-space representation of the two-mode system \cite{cui2012atom}.

In three dimensions, the bare interaction parameters read
$u_1/V = 4\pi \hbar^2 a_{\text{aa}}/(Vm_{\text{a}})$,
$u_2/V = 4\pi \hbar^2 a_{bb}/(Vm_{\text{b}})$, and
$u_3/V = 4\pi \hbar^2 a_{ab}/(Vm_{\text{ab}})$, with units of $\mathrm{J}$ \cite{similar_hamiltonian2,kohler2006production}.
Here $a_{\text{aa}}$, $a_{bb}$, and $a_{ab}$ denote atom-atom, molecule-molecule, and atom-molecule scattering lengths; $V\sim L_0^3$ with harmonic length scale, $L_0=\sqrt{\hbar/(m_{\text{a}}\Omega)}$ \cite{BEC9}, where $m_{\text{a}}$ and $\Omega$ are the ${}^{87}\mathrm{Rb}$ mass and trap frequency respectively. Number of particles in the trap: $N = 10^8$ \cite{large_N1,large_N2}. The coupling satisfies $g/\sqrt{V}=\sqrt{\mu_{\text{co}}\Delta Bu_1/V}$, yielding the effective strength $\tilde{g}=\sqrt{U_1\Delta B\mu_{\text{co}}}$. The reduced mass is $m_{\text{ab}}=m_a m_b/(m_a+m_b)$, and the binding energy is $\epsilon_b=\mu_{\text{co}}(B-B_0)$. We adopt $\Delta B=0.21$ Gauss and $B_0=1007.4$ Gauss \cite{kohler2006production}, with magnetic moment difference between closed (molecular), and open (atomic) channel is $\mu_{\text{co}}=2\mu_B$ ($\mu_B$ the Bohr magneton), and fix the detuning to $|B-B_0|=10$ Gauss. 

The coupling noise $\gamma_x$ arises from condensate-thermal collisions \cite{quantumgas2}, characterized by the collision rate $\Gamma_x=8\pi a_{\mathrm{s}}^{2}n_{\mathrm{th}}v_{\mathrm{th}}$, where $v_{\mathrm{th}}$ is the thermal velocity at temperature $T$, $8\pi a_{\mathrm{s}}^{2}$ the identical-boson scattering cross section, and $n_{\mathrm{th}}$ the thermal density associated with $N_{\mathrm{th}}$ per quantization volume. The corresponding diffusion rate is $\tilde{\Gamma}_x=8\pi^3\Gamma_x$ \cite{liu2010shapiro}.

We assume $10\%$ noise in both the coupling strength and magnetic field, giving $\gamma_x=0.1\tilde{g}$ and $\gamma_z=0.1\epsilon_b$. Even for thermal fractions approaching $30\%$, the dynamics remain condensate dominated \cite{donley2002atom}. Noise changes the response strength but leaves the intrinsic dynamics unchanged.
For convenience, the parameters governing the relaxation dynamics are summarised in the table \ref{table}.

\begin{table}
\caption{Useful parameters of the Bloch Dynamics}
 \begin{tabular}{|c|c|c|c|c|c|c|}
\hline
$U_1$ & $U_2$  & $U_3$ & $\tilde{g}$ & $\epsilon_b$ & $\gamma_x$ & $\gamma_z$ \\
\hline
$1$ & $2$ & $-1.5$ & $0.2$ & $2$ & $0.02$ & $0.2$ \\
\hline
\end{tabular}
\label{table}   
\end{table}

\section{Initialization of Bloch Dynamics with Finite Coupling Noise}\label{non zero coupling}
For the case in which only $\gamma_x$ is activated, the fluctuations of the initial Bloch-vector component velocities, $\delta v_i(0)$, together with the corresponding fluctuations of the Bloch-vector components, $\delta s_i(0)$, are given by

\begin{subequations}
 \begin{equation}
\delta v_x(0)=0, \quad \delta v_y(0)=-k \delta s_x(0)-2\sqrt{2}\tilde{g} \delta s_z(0)     
 \end{equation}
 \begin{equation}
\delta v_z(0)=-4\gamma_x \delta s_z(0)     
 \end{equation}
 \begin{equation}
 \delta \dot{v}_x(0)=k \delta v_y(0), \quad \delta \dot{v}_y(0)=-2\bigg(\sqrt{2}\tilde{g} \delta v_z(0)+2\gamma_xv_y(0)\bigg)    
 \end{equation}
 \begin{equation}
\delta \dot{v}_z(0)=2\sqrt{2}\tilde{g} \delta v_y(0)-4\gamma_x \delta v_z(0)     
 \end{equation}
\end{subequations}
Tangent of phase angles are defined as, 
\begin{subequations}
\begin{equation}
 \tan\theta_x=0,\quad\tan\theta_y=\frac{2\omega_s \delta v_y(0)}{2 \delta \dot{v}_y(0)+\gamma \delta v_y(0)}   
\end{equation}
\begin{equation}
 \tan\theta_z=\frac{2\omega_s \delta v_z(0)}{2\delta\dot{v}_z(0)+\gamma \delta v_z(0)}, \quad \tan\phi=\frac{2\omega_s}{\gamma}   
\end{equation}    
\end{subequations}
If coupling corrupted then, $\gamma=8\gamma_x$, $\tilde\omega^2=\omega^2+16\gamma^2_x$, and $\tilde{c}=4\gamma_x k^2$
\section{Specification of Initial Conditions for the Bloch Dynamics under Finite Detuning Noise}\label{non zero detuning}
In the presence of only $\gamma_z$, the fluctuations of the initial Bloch-vector component velocities, $\delta \tilde{v}_i(0)$, together with the corresponding fluctuations of the initial Bloch-vector components, $\delta s_i(0)$, are expressed as follows:

\begin{subequations}
    \begin{equation}
      \delta \tilde v_x(0)=-\frac{\gamma_z}{2} \delta s_x(0), \quad \delta\tilde v_y(0) =-k \delta s_x(0)-2\sqrt{2}\tilde{g} \delta s_z(0)  
    \end{equation}
    \begin{equation}
   \delta\tilde v_z(0)=0     
    \end{equation}
 \begin{equation}
 \delta\dot{\tilde v}_x(0)=k \delta\tilde{v}_y(0)-\frac{\gamma_z \tilde \delta v_x(0)}{2}    
 \end{equation}
 \begin{equation}
   \delta\dot{\tilde v}_y(0)=-k \tilde \delta v_x(0)-\frac{\gamma_z \tilde \delta v_y(0)}{2},\quad \delta\dot{\tilde v}_z(0)=2\sqrt{2}\tilde{g} \delta \tilde v_y(0)    
 \end{equation}
\end{subequations}
Now, tangent of phase angles are defined as, 
\begin{subequations}
    \begin{equation}
 \tan\theta_x=\frac{\omega_s \delta\tilde{v}_x(0)}{k\delta\tilde{v}_y(0)},\quad\tan\theta_y=-\frac{\omega_s \delta\tilde{v}_y(0)}{k\delta\tilde{v}_x(0)}       
    \end{equation}
    \begin{equation}
  \tan\theta_z=0       
    \end{equation}
\end{subequations}
If detuning corrupted then, we obtain 
Here, $\gamma=\gamma_z$, $\tilde\omega^2=\omega^2+\gamma^2_z/4$, and $\tilde{c}=4 \tilde{g}^2 \gamma_z$

\bibliography{bibi.bib}
\end{document}